\documentclass[12pt,a4paper]{article}
\usepackage{amssymb,amsfonts,amsmath}

\title{Randomness: quantum versus classical}

\author{Andrei Khrennikov\\
International Center for Mathematical Modelling \\
in Physics and Cognitive Sciences\\
Linnaeus University,
V\"axj\"o, SE-351 95, Sweden\\
Andrei.Khrennikov@lnu.se}

\date{}
 
\begin{document}

\maketitle

\begin{abstract} Recent tremendous development of quantum information theory led to a number of quantum technological projects, e.g., 
quantum random generators. This development stimulates a new wave of interest in quantum foundations. One of the most intriguing problems of quantum 
foundations is elaboration of a consistent and commonly accepted interpretation of quantum state. Closely related problem is clarification of the notion 
of quantum randomness and its interrelation with classical randomness. In this short review we shall discuss basics of classical theory of randomness 
(which by itself is very complex and characterized by diversity of approaches) and compare it with irreducible quantum randomness. The second part of 
this review is devoted to the information interpretation of quantum mechanics (QM) in the spirit of Zeilinger and Brukner (and QBism of  Fuchs et al.) and 
physics in general (e.g., Wheeler's ``it from bit'') as well as digital philosophy of Chaitin (with historical coupling to ideas of Leibnitz). Finally, we continue 
discussion on interrelation of quantum and classical randomness and information interpretation of QM.     
\end{abstract}

\section{Introduction}

Recently the interest to quantum foundations was rekindled by the rapid and  successful development of quantum information theory. 
One of the promising quantum information projects which can lead to real technological applications is the project on {\it quantum random generators.} Successful realization 
of this project attracted attention of the quantum community to the old and complicated problem of interrelation of quantum and classical randomness. In this short review 
we shall discuss this interrelation: classical randomness.\footnote{As based on a variety of approaches: unpredictability (von Mises), 
complexity-incompressibility (Kolmogorov, Solomonoff, Chaitin),  typicality (Martin-L\"of).} versus irreducible quantum randomness\footnote{As based on acausality  of quantum 
measurements (von Neumann) and information principles (Zeilinger and Brukner \cite{Z0}, \cite{BR1}- \cite{BR4}, \cite{Z1}, see also Kofler and Zeilinger \cite{Kofler}).} This review can be useful for researchers working in quantum information theory, both as a review 
on classical randomness and on interpretational problems of QM related to the notion of randomness. 

We emphasize the coupling between information and randomness, both in the classical and quantum frameworks. This approach is very natural in the light of modern information revolution 
in QM and physics in general. Moreover, ``digital philosophy'' (in the spirit of Chaitin) spreads widely in modern science, i.e., not only in physics, but in, e.g., computer science, artifical 
intelligence, biology. Therefore it is natural to discuss jointly randomness and information including novel possibilities to operate with quantum information and randomness outside 
of physics, e.g., in biology (molecular biology, genetics) and cognitive science and general theory of decision making \cite{UB_KHR}, \cite{BOOK}, \cite{BIO}.

\section{Random sequences: heuristics}
\label{RR4}

We start by pointing out that one has to distinguish theories of probability and randomness (although they are closely related).
At the heuristic level the difference between the notions of probability and randomness can be easily explained. First consider 
the set  $\Omega$ of all  sequences of the length $N$  composed of zeros and ones,  $x=(x_1,...,x_N), x_j=0,1.$ The uniform probability distribution  
on $\Omega$ is defined as $p(x)=1/2^N, x  \in \Omega.$ Thus all sequences are equally probable. 
Suppose, for example, that in some experiment we obtained a sequence of the form (spaces are added to make it more readable):
\[1100100100 00111111 01101010 10001000 10000101 10100011 00001000 11010011\] 
\[ 00010011 00011001 10001010 00101110  00000011 01110000 01110011 01000100 \]
\[ 10100100 00001001 00111000 00100010 00101001 10011111 00110001 11010000 \]
\[ 00001000 00101110 11111010 10011000 11101100 01001110 01101100 10001001\]
This sequence matches well our heuristics about randomness. To use it later,
we denote this sequence $x_{\rm{rand}}.$  (Here $N=258.)$ Now suppose that we obtained the sequence
of the same length of the form $x=010101.... 01$ (composed of blocks $01)$. It is clear that it cannot be treated as a random sequence. Would this sequence appear in a random experiment, it would be strange and unexpected. However, the probability 
to pick up both sequences from $\Omega$ is the same. Thus the probability calculus cannot formalize our 
heuristic image of randomness. Of course, it is clear why the second sequence is not random. The repeatability of 
the appearance of zeros and onces is a definite signal of presence of some {\it causal process}\index{causal process} producing this 
regular pattern.    But in general it is very difficult to understand whether there is some causal process behind an observed
sequence of zeros and ones. To illustrate this problem, I specially took the ``random sequence''  $x_{\rm{rand}}$ as the first 
258 digits in the binary expansion of the number $\pi$  which can be computed 
algorithmically. Thus one cannot simply proceed heuristically. Some formal theory of randomness and its interrelation with probability 
has to be developed.  

We remark that consideration of infinite sequences cannot solve the problem   of mismatching heuristics related to probability 
and randomness. The space of infinite binary sequences $\Omega$ can also be endowed with the uniform probability distribution.
Sequences which look like random and sequences which look like regular do have the same probability of appearance (in fact, zero probability).

The problem of interrelation of the notions of probability and randomness was discussed already by Laplace \cite{Laplace}: 

{\small ``We arrange in our thought, all possible events in various classes; and
we regard as extraordinary those classes which include a very small
number. In the game of heads and tails, if heads comes up a hundred
times in a row, then this appears to us extraordinary, because
the almost infinite number of combinations that can arise in a hundred
throws are divided in regular sequences, or those in which we
observe a rule that is easy to grasp, and in irregular sequences, that
are incomparably more numerous.''} 

Roughly speaking, Laplace tried to shift the problem of randomness from individual sequences to 
classes (sets) of sequences.  Although we cannot distinguish random and non-random sequences 
using probabilities of their individual appearance, we might try to estimate probability measures 
 of some classes of sequences. The set of regular sequences has a small measure;  the set of random sequences 
has an essentially larger measure. From this viewpoint, random sequences are typical and regular are atypical.   
This viewpoint to formalization of the notion of randomness led to definition of randomness  as {\it typicality}\index{typicality}
(Laplace-Ville--L\"of, see section \ref{MLOF}).  

Besides the typicality dimension of randomness Laplace also pointed to the aforementioned presence of 
a causal process as preventing a sequence to be random \cite{Laplace}: 

{\small ``The regular combinations occur more rarely only because they are
less numerous. If we seek a cause whenever we perceive symmetry, it
is not that we regard the symmetrical event as less possible than the
others, but, since this event ought to be the effect of a regular cause or
that of chance, the first of these suppositions, is more probable than
the second. On a table, we see letters arranged in this order: C o n s t a n t i n o p l e, and we judge that this arrangement is not the
result of chance, not because it is less possible than others, for if this
word were not employed in any language we would not suspect it came
from any particular cause, but this word being in use among us, it is
incomparably more probable that some person has thus arranged the
aforesaid letters than this arrangement is due to chance.''} 

This viewpoint led to formalization of the notion of an individual random sequence
through formalization of the notion of causal generation of a sequence which 
culminated in the Kolmogorov algorithmic complexity approach to randomness, \index{Kolmogorov algorithmic complexity}
sections \ref {KRA}, \ref {KRA1}. 

\section{Classical randomness}

\subsection{Approach of von Mises: randomness as unpredictability}
\label{RRR}

Von Mises (1919) theory was the first probability theory \cite{[169]}-\cite{[171]} based fundamentally on 
the principle of the  statistical stabilization of  \index{statistical stabilization} frequencies. Although this principle
was heuristically used from the very beginning of probabilistic studies, only von Mises
tried to formalize it mathematically and to establish it as one of the basic principles 
of probability theory. His theory is based on the notion of a
{\it collective} ({\it random sequence}).

Consider a random experiment $S$ and denote by $L=\{\alpha_1,..., \alpha_m\}$ the set
of all possible results of this experiment\footnote{R. von Mises did not consider probability theory as 
a purely mathematical theory. He emphasized that this is a physical theory such as, e.g., hydrodynamics. 
Therefore his starting point is a physical experiment which belongs to physics and not to mathematics.
He was criticized for mixing physics and mathematics. But he answered that there is no way to proceed with probability 
as a purely mathematical entity.}. 
The set $L$ is said to be the label set, or the set
of attributes of the experiment $S.$ We consider only finite sets $L.$ Let us consider $N$ trials  for this 
$S$ and record the results,  $x_j \in L.$  This process generates a finite sample:
\begin{equation}
\label{3.1}
x=(x_1,..., x_N\},  x_j \in L.
\end{equation}
A {\it collective} is an infinite idealization of this finite sample:
\begin{equation}
\label{3.2}
x=(x_1,..., x_N,...\},  x_j \in L,
\end{equation}
for which the following two von Mises principles are valid.

${\bf Principle \; 1}$ (statistical stabilization). This is the principle of the statistical stabilization of relative frequencies of each attribute $\alpha \in L$ 
of the experiment $S$ in the sequence (\ref{3.2}). Take the frequencies 
$
\nu_N(\alpha;x) = \frac{n_N(\alpha;x)}{N} 
$
where $\nu_N(\alpha;x)$ is the number of appearance of the attribute $\alpha$ in the first $N$ trials. By the
principle of the statistical stabilization
 {\it the frequency $\nu_N(\alpha;x)$ approaches a limit as $N$ approaches infinity, for every label $\alpha\in L.$} 
This limit 
$
P_x(\alpha)= \lim _{N \to \infty} \nu_N(\alpha;x)
$
is called {\it the probability of the attribute $\alpha$ 
of the random experiment $S.$} (Sometimes (when the  collective is fixed) 
this probability will be denoted simply as $P(\alpha).)$ 

\medskip

${\bf Principle \; 2}$ (randomness).
Heuristically it is evident that we cannot consider, for example, the sequence 
$$
x=(0, 1, 0, 1, 0, 1,...)
$$
as the output of a random  experiment. However, the principle of
the statistical stabilization holds for $x$ and $P_x(0)=P_x(1)=1/2.$ 
To consider  sequences (\ref{3.2}) as objects of probability theory,  
we have to put an additional constraint on them: {\it The limits of relative frequencies have to be stable with respect to a place selection (choice of a subsequence) in (\ref{3.2}).}
In particular, $x$ does not satisfy this principle.

However, this very natural notion (randomness) was the hidden bomb in the foundations of von
Mises' theory. The main problem was to define a class of place selections which
induces a fruitful theory. The main and very natural restriction which was set by von Mises is that a place selection
in (\ref{3.2}) cannot be based on the use of attributes of elements. For example, one cannot
consider a subsequence of (\ref{3.2}) constructed by choosing elements with the fixed label
$\alpha \in L.$  Von Mises  defined a place selection in the following way \cite{[170]}, p.9:

${\bf PS}$ {\small ``a subsequence has been derived by a place selection if the decision to retain
or reject the $n$th element of the original sequence depends on the number $n$ and
on label values $x_1,..., x_{n-1}$  of the $n-1$ preceding elements, and not on the
label value of the $n$th element or any following element''.}

Thus a place selection can be defined by a set of functions \index{place selection}
\begin{equation}
\label{psps}
f_1, f_2(x_1), f_3(x_1, x_2), f_4(x_1, x_2, x_3),..., f_n(x_1,...,x_{n-1}),...
\end{equation}
each function yielding the values 0 (rejecting the $n$th
element) or 1 (retaining the $n$th element).
Since any place selection has to produce from an infinite input sequence also an
infinite output sequence, it has also to satisfy the following restriction:
\begin{equation}
\label{pspst}
f_n(x_1,...,x_{n-1})=1 \; \mbox{for infinitely many} \;  n.  
\end{equation}

Here are some examples of place selections: 
\begin{itemize}

\item choose those $x_n$ for which $n$ is
prime; 

\item choose those $x_n$ which follow the word $01;$ 

\item toss a (different) coin;
choose $x_n$ if the $n$th toss yields heads. 

\end{itemize}

The first two selection procedures are law-like, the third selection is random. 
It is clear that all of these three procedures
are place selections: the value of $x_n$ is not used in determining whether to choose $x_n.$

The principle of randomness ensures that no strategy using a place selection rule
can select a subsequence with different odds (e.g., for gambling) than a sequence that
is selected by flipping a fair coin. Hence, the principle can be called {\it the law of excluded
gambling strategy.}  We cite Feller \cite{Feller}, pp. 198, 199: 
{\small ``The painful experience of many gamblers have taught us the lesson
that no system of betting is successful in improving the gambler
chances . . .The importance of
this statement was first recognized by von Mises, who introduced the
impossibility of a successful gambling system\index{gambling system} as a fundamental axiom.''}

Let   $x=(x_j)$ be a collective (random sequence) with the label set $L=\{0,1\},$ i.e., $x_j=0,1.$    
Given a place selection, see (\ref{psps}), (\ref{pspst}), 
let $n_1$ be the least $n$ such that $f_n(x_1,...,x_{n-1})=1,$ $n_2$ is the next such $n$, etc.  Then by the principle of randomness: 
\begin{equation}
\label{psps1}
\exists \lim_{N \to \infty}\frac{1}{N} \sum_{k=1}^N x_{n_k} = \lim_{{N \to \infty}}\frac{1}{N} \sum_{k=1}^N x_{k} =P_x(1).
\end{equation}
We stress that von Mises did not solve the problem of the existence of collectives. 

{\bf Mises-Church collectives (random sequences).} The simplest way (at least from the mathematical viewpoint) is to  proceed with special classes of 
{\it lawlike place selections.} In particular, A. Church \cite{CH} proposed to consider place selections (\ref{psps}), (\ref{pspst}) 
in which the selection functions $f_n(x_1,...,x_{n-1})$ are algorithmically computable. (We recall that $f_n$ is used to select or not the 
$n$th element of a sequence $x=(x_1,...,x_n,...).$ It is important to note that the set of all Church-like place selections
is countable, see, e.g., \cite{INT_KHR}.   The existence of 
Church's collectives is a consequence of the general result of A.  Wald \cite{[173]} which we formulate 
now.

{\bf Mises-Wald collectives (random sequences). } Let $p=(p_j=P(\alpha_j))$ be a probability
distribution on the label set $L=\{\alpha_1,..., \alpha_m\}.$ 
Denote the set of all possible sequences with elements from $L$ by the symbol $L^\infty.$ Let $\phi$ 
be a place selection. For $x \in L^\infty,$ the symbol $\phi x$ is used to denote the subsequence 
of $x$ obtained with the aid of this place selection.
Let $U$ be some family of place selections. We set
$
X(U; p)= \{x \in L^\infty: \forall \phi \in U \; \lim_{N \to \infty} \nu_n(\alpha_j; \phi x)=p_j, j=1,...,m\},
$
where as usual $\nu_N(\alpha;y), \alpha \in L,$ denotes the relative frequency of the appearance of the label $\alpha$ 
among the first $N$ elements of the sequence $y \in L^\infty.$ 

{\bf Theorem 1.}  (Wald \cite{[173]}). {\it For any countable set $U$ of place selections and any probability
distribution $p$ on the label set $L,$ the set of sequences $X(U; p)$ has the cardinality of
the continuum.}

\medskip

By Wald's theorem\index{Wald theorem} for any countable set of place selections $U$ the frequency theory
of probability can be developed at the mathematical level of rigorousness. R. von
Mises was completely satisfied by this situation (see \cite{[171]}). 

\medskip

{\bf Ville's attack to von Mises theory}

However, a new cloud appeared on the sky. This was the famous  
{\it Ville's objection}\index{Ville objection} \cite{Ville}.

{\bf Theorem 2.} (Ville). {\it Let $L=\{0,1\}$ and let $U=\{\phi_n\}$ be a countable set
of place selections. Then there exists $x\in  L^\infty$  such that
\begin{itemize}

\item  for all $n,$ 
$$
\lim_{N\to \infty} \sum_{j=1}^N (\phi_n x)_j= 1/2;
$$

\item for all $N,$ 
$$
\sum_{j=1}^N (\phi_n x)_j \geq 1/2.
$$ 
\end{itemize}
}

Such a sequence $x$ of zeros and ones is a collective with respect to $U,$ $x \in X(U;1/2),$ but seems to be far too
regular to be called random. At the same time from the measure-theoretic viewpoint  the existence of such sequences is not a problem. The set of such sequences 
has the Lebesgue measure zero. (We recall that any sequence of zeros and ones can be identified with a real number from the segment 
$[0,1]).$ Here we see the difference between the treatment of randomness as unpredictability (a la von Mises) and as typicality (see section 
\ref{MLOF} for the latter -- theory of Martin-L\"of). 
   
\subsection{Laplace-Ville-Martin-L\"of: randomness as typicality} 
\label{MLOF}

Ville  used Theorem 2 to argue that collectives in the sense of von Mises
and Wald do not necessarily satisfy all intuitively required properties of randomness.
J. Ville introduced  \cite{Ville} a new way of characterizing random sequences (cf. with Laplace \cite{Laplace}), based on the following
principle:

{\bf Ville's Principle:}\index{Ville Principle} {\it A  random sequence should satisfy all properties of probability one.}\footnote{Of course, Ville's approach matches Laplace viewpoint \cite{{Laplace}}
on randomness as typicality. }

Each property is considered as determining a {\it  test of randomness.}\index{randomness test} According to Ville \cite{Ville}, a sequence 
can be considered random if it passes all possible tests for randomness.

However, this is impossible: we have to choose countably many of
those properties (otherwise the intersection of the uncountable family of sets of probability 1 can have probability less than 1 or simply be 
nonmeasurable; in the latter case the probability is not defined at all). Countable families of properties (tests for randomness) 
can be selected in various ways. A random sequence passing one countable sequence of tests can be rejected by another. This brings ambiguity in the Ville approach to randomness as typicality (i.e., holding some property with probability one).

It must be underlined that the Ville principle is really completely foreign to
von Mises. For von Mises, a collective $x\in  L^\infty$  induces a probability on the set of labels
$L,$ not on the set of all sequences $L^\infty.$ Hence, for von Mises (and other scientists interpreting randomness as unpredictability\index{unpredictability} 
in an individual sequence),  there is no connection  between
properties of probability one in $L^\infty$ and properties of an individual collective.

Later (in 1970th) P. Martin-L\"of (motivated by the talk of A. N. Kolmogorov at 
the Moscow Probability Seminar) \cite{[142]} 
solved the problem of ambiguity of the Ville interpretation of randomness as typicality.\index{typicality} He
proposed to consider {\it recursive (algorithmic) properties of probability
one}, i.e., the properties which can be tested with the aid of algorithms. Such an
approach induces the fruitful theory of recursive (algorithmic) tests for \index{recursive test} randomness (see, for example,
\cite{[143]}, \cite{[144]}). The key point of this ``algorithmic tests'' approach to the notion of randomness is that it is possible 
to prove that there exists the {\it universal algorithmic test}.  A sequence is considered random if it passes this 
universal test. Thus the class of typicality-random sequences is well defined. Unfortunately, this universal test\index{universal test} of randomness
cannot be constructed algorithmically (although by itself it is an algorithmic process). Therefore, although we have the well 
defined class of Martin-L\"of random sequences, we do not know what the universal test for such randomness looks like. Hence, for a concrete 
sequence of trials we cannot check algorithmically whether it is random or not -- although we know that it is possible to perform
such algorithmic check.

\section{Kolmogorov: Randomness as Complexity}
\label{KRA}

It is well know that personally A. N. Kolmogorov was not satisfied by his own measure-theoretic approach 
to probability (private communications of his former students). He sympathized to the von Mises approach 
to probability in which randomness was no less fundamental than probability. In 1960th he turned again to the 
foundations of probability and randomness and tried to find foundations of randomness by reducing this notion to
the notion of complexity, 1963 \cite{[134]a}-\cite{[135]}.   
Thus in short the Kolmogorov approach can be characterized as {\it randomness as complexity.}\index{Kolmogorov complexity}
Another basic point of his approach is that complexity of a sequence has to be checked {\it algorithmically.}

Let $L=\{0,1\}.$ Denote by $L^*$ the set of all {\it finite sequences} (words)  in the alphabet $L.$ 

\medskip

{\bf Definition 1} (Kolmogorov). {\it Let $A$ be an arbitrary algorithm. The algorithmic complexity of a
word $x$ with respect to $A$ is $K_A(x)=  \min l(\pi),$ where $\{\pi\}$ are the programs which
are able to realize the word $x$ with the aid of $A.$}

\medskip

Here $l(\pi)$ denotes the length of a program $\pi.$ This definition depends on the structure
of the algorithm $A.$ Later Kolmogorov proved the following fundamental theorem:

\medskip

{\bf Theorem 3.}  (Kolmogorov,  Solomonoff)\footnote{Kolmogorov published this result in 1965 {[134]}. In fact, the first proof was presented in Solomonoff's preprint  \cite{Sol0}
 in 1960. However, practically nobody paid attention to 
this work. Its importance was not recognized by the scientific community. When Kolmogorov became aware of this work, he openly recognized the priority of Solomonoff. In fact, this Kolmogorov's
recognition played the crucial role in advertising research of Solomonoff who first became famous in Soviet Union and only later in Western countries.
 Kolmogorov's contribution to establishing  this area of research was memorized in assigning the name {\it Kolmogorov complexity} to the 
algorithmic complexity.This does not diminish the role of  the contribution of Solomonoff \cite{Sol0}, \cite{Sol} 
(and Chaitin \cite{Chaitin}).} {\it There exists an algorithm $A_0$ (optimal algorithm) such that, for any 
algorithm $A,$ there exists a constant} $C>0,$
\begin{equation}
\label{HH}
K_{A_0}(x) \leq  K_{A}(x) +C .
\end{equation}

\medskip

It has to be pointed out that optimal algorithm is not unique. 

The complexity $K(x)$  of the word $x$ is by definition equal to the complexity $K_{A_0}(x)$ 
with respect to one fixed  optimal algorithm $A_0.$

The original idea of Kolmogorov  \cite{[134]a}-\cite{[135]} was that complexity $K(x_{1:n})$  of the initial
segments $x_{1:n}$  of a random sequence $x$ has to have the asymptotic 
$\sim n$
\begin{equation}
\label{HHa}
K(x_{1:n}) \sim n , n \to \infty,
\end{equation}
i.e., we might not find a short code for $x_{1:n}.$ 

However, this heuristically attractive idea was rejected as a consequence of  the   objection of  Martin-L\"of \cite{[144]}. 
To discuss this objection and connection of the Kolmogorov complexity-randomness 
with Martin-L\"of typicality-randomness, we proceed with
{\it conditional algorithmic complexity} $K(x; n),$ instead of complexity $K(x).$ Conditional complexity
$K(x; n)$ is defined as the length of the minimal program $\pi$ which produces the output
$x$ on the basis of information that the length of the output $x$ is equal to $n.$  

{\bf Theorem 4.} (Martin-L\"of) {\it   For every binary sequence $x,$
\begin{equation}
\label{HHb}
K(x_{1:n};n) < n -\log_2 n,
\end{equation}
for infinitely many n.}

Hence, {\it Kolmogorov random sequences, in the sense of the definition (\ref{HHa}),
do not exist.}

Another problem of the Kolmogorov approach to randomness as algorithmic complexity is that we ``do not know'' any
optimal algorithm $A_0,$ i.e., {\it the Kolmogorov complexity is not algorithmically computable!} However, the latter 
problem can be partially fixed, because there exist algorithmic methods to estimate 
this complexity (from above and from below).

\subsection{Kolmogorov-Chaitin Randomness} 
\label {KRA1}

As we have seen,  in its straightforward form the Kolmogorov algorithmic complexity does not lead to a nontrivial notion of a random sequence. 
However, as it happens \cite{Chaitin}, a fruitful theory of individual random sequences is very near, it is enough to slightly  modify
the notion of Kolmogorov complexity. \index{Kolmogorov-Chaitin randomness}

First, we remark that the notion of algorithm can be formalized as a {\it partial computable (recursive) function} \index{partial computable function}
$A: L^* \to L^*$ (here ``partial'' means that in general such a function is  defined only on some subset $D=D_{A}$ of  $L^*$). 
Thus Kolmogorov complexity of a word $x \in  L^*$ with respect to $A$ equals the length of the shortest 
word $\pi\in D$ such that $A(\pi)=x: K_{A}(x) = l(\pi).$   If such  $\pi\in D$ does not exist, then
 $K_{A}(x)= + \infty.$ 

A prefix of a word $x = x_1 \dots x_n$ is a  $\widehat x = x_1 \dots x_{m},$ where $m \leq n.$
A subset $D$ of  $L^*$ is called {\it prefix free} if no word in  $D$ is a prefix of another member of $D.$
A real world example of a prefix free set (over the alphabet of decimal digits)  is the set of country dialing 
codes in the international telephone system.\index{prefix free complexity}

\medskip

{\bf Definition 2.}  {\it Let $A$ be an arbitrary algorithm, a partial computable function, 
with prefix free domain of definition $D.$  Algorithmic prefix free complexity of a 
word $x \in  L^*$ with respect to $A$ equals to the length of the shortest 
word $\pi \in D$ such that $A(\pi)=x: \tilde{K}_{A}(x) = l(\pi) .$   If such  $\pi\in D$ does not exist, then
 $\tilde{K}_{A}(x)= + \infty.$ }

\medskip

{\bf Theorem 5.} {\it There exists an optimal  prefix free algorithm (partial computable function) $A_0$  such that, for any 
prefix free algorithm $A,$ there exists a constant} $C>0,$
\begin{equation}
\label{HH}
\tilde{K}_{A_0}(x) \leq  \tilde{K}_{A}(x) +C .
\end{equation}

Prefix free complexity $\tilde{K}(x)$  of the word $x$ is (by definition) equal to the complexity $K_{A_0}(x)$ 
with respect to one fixed (for all considerations) optimal prefix free algorithm $A_0.$\index{Kolmogorov-Chaitin randomness}\index{incompressible}

\medskip

{\bf Definition 3.} {\it A sequence $x \in L^\infty, L=\{0,1\}$ is called Kolmogorov-Chaitin random if it is 
incompressible (no initial segment 
of $x$ can be compressed more than for a fixed finite number of bits) or in other words:}
\begin{equation}
\label{hhyyuu}
\exists b>0:  \tilde{K}(x_{1:n}) \geq n-b \; \mbox{for all} \;n.
\end{equation}

\medskip

{\bf Equivalence of Kolmogorov-Chaitin and Martin-L\"of  approaches to randomness}

At the first sight the Kolmogorov-Chaitin and Martin-L\"of  approaches to randomness differ crucially. The first one is about randomness 
of an individual  sequence. There is no reference to other sequences; one is not interested in how typical is this concrete sequence in an
ensemble of all possible sequences. We can say that Kolmogorov-Chaitin randomness is determined intrinsically  and 
 Martin-L\"of  randomness is  determined externally.  Therefore the following result is really surprising:
 
 \medskip
 
 {\bf Theorem 6.} (Schnorr \cite{[158]}) {\it A  sequence is  Martin-L\"of  random if and only if its Kolmogorov-Chaitin random.}\index{Kolmogorov-Chaitin randomness}
\index{Martin-L\"of  randomness}
 
 \medskip
 
 {\bf Coupling between  Kolmogorov-Chaitin, Martin-L\"of  and Mises-Wald-Church  randomnesses}\index{Mises-Wald-Church  randomnesses} 
 
{\bf Theorem 7.} (Invariance of randomness with respect to place selections) 
{\it Let   $x=(x_j)$ be a Kolmogorov-Chaitin (Martin-L\"of) random sequence.     
Given an algorithmically computable  place selection, see (\ref{psps}), (\ref{pspst}),
 let $n_1$ be the least $n$ such that $f_n(x_1,...,x_{n-1})=1,$
$n_2$ be the next such $n$, etc. Then $x=(x_{n_k})$ is also a Kolmogorov-Chaitin (Martin-L\"of) 
random sequence.}

\medskip

It is also possible to prove that any Kolmogorov-Chaitin (Martin-L\"of) random sequence satisfies the principle of statistical stabilization, i.e., relative frequencies tend 
to limits - probabilities. Finally, we obtain the following  important result providing a partial connection with  von Mises' notion of collective (with Church's flavor):

\medskip

{\bf Corollary 1.} {\it Any Kolmogorov-Chaitin (Martin-L\"of) random sequence is also Mises-Wald-Church random.}

Of course, the inverse is not correct: as was shown by Ville, there exist Mises-Wald-Church random sequences which 
do not pass the basic statistical tests.

In the community working on foundations of randomness 
the assertion that {\it Martin-L\"of randomness or equivalently Kolmogorov-Chaitin
randomness captures the ``true notion of randomness''' conforming to our intuition
is sometimes called the Martin-L\"of-Chaitin thesis.} (The Martin-L\"of-Chaitin thesis,\index{Martin-L\"of-Chaitin thesis} \index{Church-Turing thesis}
like the Church-Turing thesis for the definition of algorithm, is not a mathematical
proposition that can be proved or refuted.) Thus in this community the problem 
of formalization of the intuitive notion of randomness is considered to be solved.
Personally I do not think so.  

Of course, the realization of the Kolmogorov-Solomonoff-Chaitin 
program  on randomness as complexity-incompressibility was one of the most important 
contributions to theory of randomness. Now we have a rigorous mathematical theory 
of individual random sequences. In the same way the Martin-L\"of theory of algorithmic 
statistical tests provides a rigorous mathematical description of randomness as typicality.
This theory establishes the solid theoretical foundation for the widely applicable method 
of testing of (pseudo)random generators, including the NIST test. \index{NIST test} An output of a 
(pseudo)random generator has to pass a block of algorithmically designed tests 
to be ``recognized'' as a (pseudo)random sequence. Schnorr's proof that the 
Kolmogorov-Solomonoff-Chaitin  and Martin-L\"of approaches match perfectly 
can be considered as culmination of development of theory of randomness in 
20th century.

Starting with 1960th (and even earlier with the works of Turing) 
the evolution of mathematical theory of randomness was closely connected to the computer revolution.
Development of the art of programming excited people about the idea of algorithmic computability. 
In the light of this revolution, it was fashionable to formulate the problem of randomness with the aid of language of computability. However, nowadays
when  the use of advanced computer programs became the everyday 
routine and people are not blindfolded by the light of programming anymore, one can ask honestly whether the whole project 
of the algorithmically based randomness was really so much justified. The dream about creation of computer-like 
artificial intelligence evaporated. Mind seems not to be  driven by computer programs. R. Penrose \cite{P1} 
rightly pointed to the transcendental nature of human mind. In this new context one can try to start a new project of formalization 
of randomness which is not based on algorithmic computability.

The situation is worse for the approach originated by von Mises (while its importance was recognized already 
by Laplace) and based on the interpretation of randomness as unpredictability. On one hand, the mathematically
rigorous formalization of von Mises notion of collective suffers of Ville's objection. On the other hand, in  the 
 Kolmogorov-Solomonoff-Chaitin-Martin-L\"of framework one can be only sure that a random sequence is 
unpredictable in the sense of Mises-Church, but not vice versa.

We remark that randomness as unpredictability  seems to be the most important for applications. In principle, 
user is not interested explicitly in complexity of an output of  a (pseudo)random generator or whether this output 
passes a block of algorithmic tests. (We remark that the sequence $x_{\rm{rand}}$ from section \ref{RR4} composed of the digits of the number $\pi$ 
passes the majority of standard tests of randomness. A special test had to be developed, the so-called $\pi$-test, in order to block 
usage of this sequence.) 
We are interested only in a possibility to predict the output or, at least, to find some patterns
in it. However, as we have seen, the direct formalization of randomness-unpredictability  has not yet been created.
The von Mises approach (or something totally novel) still waits for its time to come.    
We remark (private communication of A. Shiryaev) 
that Kolmogorov died with the hope that in future  a new and unexpected approach to the notion of randomness 
will be elaborated. And, of course, mathematicians will continue to work on this problem. However, it may happen that  future 
attempts will never lead to a mathematically acceptable notion of randomness.  This was the final point of
my lectures given  at IQOQI (Vienna) in May-June 2014. 
In the after-talk discussion various opinion were presented; in particular, prof. Zeilinger conjectured that such a painful
process of elaboration of the mathematical theory of randomness is simply a consequence of the methodological mistreatment of this notion.
It might be that {\it randomness is not mathematical, but a physical notion.} Thus one has to elaborate physical procedures guarantying randomness
of experimentally produced sequences and not simply try to construct such procedures mathematically. 
In some sense Zeilinger's proposal  is consonant with von Mises' proposal: {\it to find a collective, one simply has to go to a casino.}
Zeilinger proposes to go to his quantum optics lab at Boltzmanngasse 3.   

\section{Irreducible Quantum Randomness from Violation  of Causality}
\label{VN_RD}

In his fundamental monograph \cite{VN}   von Neumann pointed out that quantum randomness 
is individual, e.g., even an individual electron is intrinsically random, while classical randomness is related to variation of states 
in an ensemble, i.e., it is reducible to such ensemble variations. 
 In particular, von Neumann remarked \cite{VN}, pp. 301-302, that, for measurement of some quantity $R$ for an ensemble of systems (of any origin), 

{\small ``it is not surprising  that $R$ does not have a sharp value
..., and that a positive dispersion exists. However, two different reasons for this behavior a priori conceivable:

1. The individual systems $S_1,..., S_N$ of our ensemble can be in different states, so that the ensemble  $[S_1,..., S_N]$ is defined by their relative frequencies.
The fact that we do not obtain sharp values for the physical quantities in this case is caused by our lack of information: we do not know in which state we are measuring,
and therefore we cannot predict the results.

2. All individual systems $S_1,..., S_N$ are in the same state, but the laws of nature are not causal. Then in the cause of the dispersion is not our 
lack of information, but is nature itself, which has disregarded the principle of sufficient cause.''}

Thus, for him, quantum randomness is statistical exhibition of violation of causality, violation of  the {\it principle of sufficient cause.} We now compare this kind of randomness with classical 
interpretations of randomness, see section \ref{randomness}: 

\begin{enumerate}
\item unpredictability (von Mises), \index{unpredictability}
\item complexity-incompressibility (Kolmogorov, Solomonoff, Chaitin),\index{complexity-incompressibility }
\item typicality (Martin-L\"of).\index{typicality}
\end{enumerate}

We start with Kolmogorov algorithmic complexity. \index{Kolmogorov algorithmic complexity} One can argue that violation of the principle of sufficient cause 
has to  imply impossibility of nontrivial compression of information in a string
of bits produced by measurements of a quantum observable $A$ for an ensemble of systems prepared in the identical state $\psi,$ - that is, impossibility to write a program which would
produce this sequence and be essentially shorter than the sequence. Really, it is difficult to imagine how any type of algorithmic process (different 
from simple output of the result of measurement)  can be represented mathematically in the absence of the classical state representation 
which would encode possible outputs of measurement.\footnote{For example, in classical statistical mechanics by determining (precisely) the state 
of a particles, by the phase-space point $(q,p),$ we determine (precisely) its energy, $E= p^2/2m +V(q).$  If the potential $V(q)$ is a computable function, e.g., a polynomial with rational 
coefficients, the energy-value can be computed algorithmically.} However, it is not easy to formalize this kind 
of reasoning.  

Thus if the principle of sufficient cause is really violated for quantum systems  and one can really connect its violation to impossibility to shorten representation of the sequence of results 
of measurement, then the output sequences of quantum measurements must have high Kolmogorov complexity and quantum measurements can be used for creation of random numbers
(in the sense of  Kolmogorov's complexity approach).\footnote{We remark that surprisingly nobody tried to estimate  Kolmogorov complexity for outputs of quantum measurements, e.g., 
for Bell's test. We remark that, although the Kolmogorov measure of  complexity is incomputable, it can be effectively estimated.  Estimation of complexity of outputs of quantum measurements 
is an exciting project which may clarify a lot in interrelation of classical and quantum randomness.}
This reasoning provides,  in fact, the foundational (philosophical) basis for the project on {\it quantum random generators} \cite{77}. 

In the light of the previous consideration it seems that  justification of proper functioning of quantum random generators can be done only from physical principles. 
Surprisingly mathematics still plays a crucial role,
 since it is heavily exploited in so-called no-go theorems saying that it is impossible to introduce hidden variables, parameters which provide a finer description of the state of a 
quantum system than given by its quantum state.  Violation of  the principle of sufficient cause is incompatible with existence of subquantum models with hidden variables.
Roughly speaking, one cannot start selling quantum random generators before a loophole free test rejecting existence of hidden variables is successfully performed.\footnote{Recently 
experimenters performed  a few tests, e.g.,  \cite{Hensen}, \cite{Giustina}, \cite{Shalm} which claim to be loophole free. However, it is too early to declare that the problem of loopholes was completely solved.
The experimental data should pass independent statistical analysis, cf. \cite{ST1}, \cite{ST2}.}  

Von Neumann understood very well the role of no-go theorems in justification of his thesis about violation of causality by quantum systems and he formulated and presented \cite{VN}
a sketch of proof of the first no-go theorem, nowadays known as the {\it von Neumann theorem.} However, his theorem was criticized, e.g., by Bell \cite{B0}, \cite{B} 
 and Ballentine \cite{[13]}, \cite{[15]} as based on unphysical assumptions
about the rules for coupling an imaginable classical hidden variable model of quantum phenomena and the genuine quantum model.  
This theorem is considered as having no physical impact. 

Now the most widely discussed  no-go theorem is due to Bell \cite{B}. However, as well as the von Neumann theorem, it is based on concrete rules 
coupling imaginable classical hidden variable model and the genuine quantum model \cite{KHR_CONT}. Adequateness of these rules to real physical situations also can be criticized   
\cite{DBB},  \cite{INT0}, \cite{INT_KHR},  
\cite{KHR_CONT}, \cite{KP1}, \cite{KP2},  \cite{Hess1}, \cite{Hess2}, \cite{Raedt1}, \cite{MIH},  \cite{Theo}. This is, in fact, the main problem of all 
no-go theorems \cite{KHR_CONT} as attempts to reject all possibly imaginable models with hidden variables
and imaginable rules coupling them with the quantum model. Even Bell  pointed 
out  \cite{B7} {\small ``that what is proved, by impossibility proofs, is lack of imagination.''} (We emphasize that the presented consideration does not question the recent experimental success
in performing the final loophole free experiment \cite{Hensen}, \cite{Giustina}, \cite{Shalm}.)

Let us assume that Bell's theorem really can be considered as describing physically reasonable coupling between the most general model with hidden variables 
and QM (as the majority  of the quantum community believes). There is still one fundamental problem preventing justification of von Neumann's statement about violation of    
the principle of sufficient cause.
Bell's theorem rejects only local hidden variable models,  i.e., models preventing faster than light communications. Thus von Neumann was right and 
the quantum random generators really produce random sequences (at least in the framework of  Kolmogorov's algorithmic complexity) only if nature were not too exotic, i.e., 
superluminal communications were impossible.  

\medskip

As we know (section{randomness}), the Kolmogorov and Martin-L\"of approaches to randomness lead to the same class of random sequences. Therefore one can proceed formally and say 
that above reasoning that quantum randomness implies Kolmogorov's randomness also leads to the implication:    quantum randomness$\mapsto$ Martin-L\"of randomness.
Thus a sequence produced by a quantum random generator has to pass the universal Martin-L\"of test and, in particular, any finite block of algorithmic tests, e.g., the NIST test.
Such type of reasoning is very popular in the community working with quantum random generators. It seems that one needs not to check whether the output of a quantum random 
generator would pass, e.g., the NIST test. There is an objection to such a viewpoint. It  is based on recognition that each quantum experiment depends on numerous
``technicalities''  modifying the output. That is, the actual output may essentially differ from output expected from the theoretical analysis of the experimental design. Thus, in any event,
the NIST test is needed to certify a quantum random  generator. 

\medskip 

Now we turn to the notion of randomness as unpredictability, a la von Mises. We repeat that  
von Mises' principle of randomness  can be treated as  {\it the law of excluded
gambling strategy.} However, such a strategy definitely does not exist if  the principle of sufficient cause is violated. 
Hence, under the latter assumption outputs of quantum measurements 
can be considered as random, from the von Mises viewpoint, i.e., as unpredictable.  

We point out that violation of the principle of sufficient cause is state dependent. If the state $\psi$ of a system is one of the eigenstates of the operator $A$ 
representing a quantum observable, then we can predict the result of measurement with probability one. Thus this principle has to be used with caution.   

 \medskip 
 
All previous considerations were devoted to matching of the notion of quantum randomness to the standard notions elaborated in mathematics. As we have seen, by 
assuming that a sequence is produced by a quantum source of randomness one can be sure that it is random in the classical sense. Thus to be random in the classical sense 
is a necessary condition of quantum randomness. Is it sufficient? The canonical answer is ``no''. 
It is typically claimed that only quantum randomness is genuine randomness.

\section{Lawless Universe? Digital Philosophy? }

{\it Where did complexity of the Universe come from?}
This is one of the most fundamental problems of modern science. One of the first scientists who took this problem seriously was Leibniz, see 
Chaitin \cite{Chaitin1} for 
the excellent presentation of Leibniz's views on this problem. \footnote{This book is the apotheosis of  the algorithmic approach, 
not only to randomness and complexity, but to science in general.  (Of course, it is surprising that in this book Kolmogorov  was not mentioned at all!)}   
And to explain the origin of Universe's complexity Leibniz ``simply'' appealed to God.
  
Leibniz was interested in distinguishing  lawful and lawless experimental data. He presented 
a beautiful example illustrating this problem. He proposed us to scatter points at random on a sheet of paper,  closing the eyes and stabbing at the paper with a pen many times,
say a few hundreds. The output will be a randomly looking pattern on the sheet. However, Leibniz pointed out that even for this data one can easy find a mathematical law,
in fact, a polynomial curve, that passes through all these randomly chosen points. To show this he used the concrete application of Lagrangian interpolation 
procedure. 

For reader's convenience, we recall that the Lagrange interpolating polynomial is the polynomial $P(x)$ of degree $\leq (n-1)$ 
that passes through the $n$ points $(x_1,y_1=f(x_1)), (x_2,y_2=f(x_2)), ..., (x_n,y_n=f(x_n)),$ and is given by
\begin{equation}
\label{LLKK}
P(x)	=	\frac{(x-x_2)(x-x_3)...(x-x_n)}{(x_1-x_2)(x_1-x_3)...(x_1-x_n)} y_1+
\end{equation}
$$
\frac{(x-x_1)(x-x_3)...(x-x_n)}{(x_2-x_1)(x_2-x_3)...(x_2-x_n)} y_2+...
$$
$$
+ \frac{(x-x_1)(x-x_2)...(x-x_(n-1))}{(x_n-x_1)(x_n-x_2)...(x_n-x_{n-1})} y_n.	
$$

Thus, in spite of the fact that the generation of the aforementioned pattern satisfies the heuristic criteria of randomness as unpredictability, we cannot say that the output pattern 
is lawless. It seems that Leibniz was the first who framed the problem of distinguishing lawfulness and lawlessness correctly: 
not as distinguishing between lawful and totally lawless processes, but between processes having different 
degree of complexity.   We see that the complexity of the mathematical law (\ref{LLKK}) rapidly increases with the increase of the size of the pattern.  
This Leibniz  reasoning was the first step towards the modern theory of complexity and randomness.  Following Leibniz, we can say that the basic 
task of a scientist is not just to find mathematical laws describing natural (or mental) phenomena, but the simple laws, laws of low complexity, which at the same time produce
sufficiently rich patterns of data (to be of some interest for scientists). Again from the above toy example, one can see that the ``majority of mathematical laws'' are complex, 
the appearance of a simple law with rich output is merely an exception. And, for Leibniz, such exceptionally simple and fruitful laws could appear only in accordance with God's 
plan: {\small ``God has chosen that which is the most perfect, that is to say, in which at the same time the hypotheses are as simple as possible, and phenomena are as rich as possible.''}   
Here Leibniz was cited again by following Chaitin \cite{Chaitin1} who used this citation for the following passage:

{\small``The complexity of the Universe is combined from the complexity of laws for expressing of natural phenomena and the complexity of initial and boundary conditions. 
If initially the Universe was described by simple initial-boundary conditions (as by the Big Bang scenario), then the complexity of the Universe is due to complexity 
of laws for expression of natural phenomena. Thus the complexity of the Universe can be identified with complexity of its laws and the latter has to be measured   
as the algorithmic complexity''.}

\medskip

However this way of thinking cannot explain where this complexity of laws came from. Although Leibniz's views are very sympathetic for Chaitin, he is not consistent enough to 
appeal to God. Instead of such an appeal to God's plan of creation of the perfect world, he referred to {\it quantum randomness} as generating patterns which cannot be described by simple 
laws.  However, in the purely classical considerations his reference to quantum is really illogical.
How can one refer to quantum randomness if the usual classical coin tossing generates a pattern which is algorithmically so complex as a pattern produced by a quantum random 
generator? We cite book \cite{Chaitin1} again, p. 119:

{\small ``This idea of an infinite series of independent tosses of a fair coin  may sound like a simple idea, a toy physical model, but it is a serious challenge, indeed a horrible nightmare, for any attempt 
to formulate a rational world view! Because each outcome is in fact that is true for  {\bf no reason}, that is true only by accident!''}

Chaitin can be really considered as one of the fathers of {\it digital philosophy}\index{digital philosophy}  which, in particular, led to {\it digital physics}\index{digital physics} culminating  in
 Wheeler's statement \cite{Wheeler}: 

{\small ``It from bit. Otherwise put, every 'it' every particle, every field of force, even the space-time continuum itself derives its function, its meaning, its very existence entirely -
even if in some contexts indirectly from the apparatus-elicited answers to yes-or-no questions, binary choices, bits. 'It from bit' symbolizes the idea that every item of the
 physical world has at bottom - a very deep bottom, in most instances - an immaterial source and explanation; that which we call reality arises in the last analysis from 
the posing of yes-no questions and the registering of equipment-evoked responses; in short, that all things physical are information-theoretic in origin and that this is a participatory universe.''}

In this paper  we are mainly interested in quantum theoretical version of informational physics, as Zeilinger-Brukner informational interpretation \cite{Z0}, \cite{BR1}- \cite{BR4}, \cite{Z1}
(see also Kofler and Zeilinger \cite{Kofler}) and QBism of Fuchs et al. \cite{Caves1}, \cite{Caves2}, \cite{Fuchs2a}- \cite{Fuchs6}. 
These approaches, while being a part of information physics, do not match the digital philosophy precisely. Here (in QM) information 
is considered as a primary physical quantity  which cannot be defined in terms of other more fundamental variables. Opposite to Chaitin and Wheeler, they (Zeilinger et al. and Fuchs et al.) 
put the transcendental content in the notion of information which matches perfectly with the transcendental content of a quantum state. For them information need not be produced 
algorithmically. Opposite to Chaitin, they are not afraid to use real and complex numbers. It seems that complex numbers really represent the physical content of phenomena.
(I am not completely sure, but it seems that QM would not work with algorithmically computable complex amplitudes.)       

We also mention the information  viewpoint on Bohmian mechanics based on the {\it active information} interpretation of the quantum potential. This interpretation was elaborated by Bohm 
and Hiley \cite{[26]}. It is amazing that even an ontological model of quantum phenomena, Bohmian mechanics, naturally generates the purely information interpretation of its basic
entity, the quantum potential.\footnote{The active information interpretation opened the door for applications of the formalism 
of Bohmian mechanics outside physics, in particular, in mathematical 
modeling of quantum-like cognition \cite{[26]}, later this formalism was explored in behavioral finances  \cite{INT_KHR}, \cite{UB_KHR}.} 

 \section{Unpredictability and Indeterminism} 
\label{UNP}

Unpredictability is very often coupled with indeterminism. The latter is the impossibility to describe generation of data by a  dynamical map:
\begin{equation}
\label{LLKK1}
y=U(x_0),
\end{equation}
where $U$ is a map from the input $x_0$ to the output $x.$

In the simplified picture of random processes determinism implies predictability, so no randomness. To be unpredictable
a process has to be indeterministic. However, this picture does not match the real situation. 

Consider the basic example of a collective, Mises random sequence: an infinite series of independent tosses of a fair coin.    It is also random from the viewpoint of Kolmogorov, i.e., 
in the framework of algorithmic complexity. Hence, it is Martin-L\"of random. However, a coin is a classical mechanical system, and its motion is described by Newtonian mechanics, 
\cite{CT1}, \cite{CT2}, \cite{CT3}, \cite{COIN}. 
Hence, one can construct the corresponding dynamical map  (\ref{LLKK1}). Thus if we know the initial condition, we can predict the outcome of a coin toss. The process of generation
of this (Mises-Kolmogorov-Martin-L\"of) random sequence is totally deterministic. Its unpredictability is just a matter of imprecision in determination of initial conditions.  Nowadays this
trade between (un)predictability and (im)precision in determination of $x_0$ can be numerically modeled \cite{COIN}. The latter paper contains a detailed mechanical model of the coin tossing dynamics. 
The results of the corresponding numerical simulation were presented graphically.  It was shown that if the imprecision in selection of $x_0$ is less than $\epsilon,$ where 
$\epsilon$ depends on parameters of the model (see \cite{COIN} for details), then one can predict the outcome of each coin toss. However, if one can determine $x_0$ with accuracy only up to 
a ball of some radius larger than this   $\epsilon,$ then the outcome cannot be determined in advance. Thus the story about coin tossing is a story about the precision of determination of 
initial conditions.    Since this is one of the basic examples of  the Mises-Kolmogorov-Martin-L\"of random sequences, we conclude that the modern mathematical theory of randomness 
does not contradict determinism in sequence generation. That is why Chaitin \cite{Chaitin1} has to refer to quantum randomness to emphasize the lawlessness dimension of randomness.
To be more precise, we have to speak about the complexity dimension. However, the example of coin tossing shows that there is nothing about complexity of physical laws. 
The dynamical  equations \cite{COIN}  are simple Newtonian equations. At the same time Kolmogorov complexity of the output is very high. 
What does this mean? Simply that {\it Kolmogorov complexity 
is not an adequate measure of complexity of physical laws behind generation of sequences of outputs.}

What is the main problem in matching the Kolmogorov approach and physics? This is consideration of solely algorithmically 
representable  laws. The algorithmic-computability approach well serves the purposes of computer science and artificial intelligence, but 
not physics. All basic physical models contain some {\it transcendental}\index{transcendental} element. For example, Newtonian mechanics is based on {\it real numbers.} \index{number, real}
We remind a few measure-theoretic facts about reals. 
\begin{itemize}
\item Consider the segment $[0,1]$ and  probability $p_L$ given by the linear Lebesgue measure; here $p_L([a,b])=b-a.$ Then probability that a randomly selected number from $[0,1]$ is rational \index{number, rational} equals to zero, the same is valid for algebraic numbers\index{number, algebraic} (solutions 
of algebraic equations). Thus probability to get a transcendental number\index{number, transcendental} is one. 
\item One can introduce the notion of a {\it computable real number}. \index{number, computable} Real numbers 
with probability one are noncomputable. 
\end{itemize}

The classical model of natural phenomena is fundamentally noncomputable. The main problem of digital philosophy 
and digital physics is that they try to identify the human brain with  computer. The latter definitely can operate only with computable quantities, but the former 
 can easily make transcendental\index{transcendental} steps in reasoning, see R. Penrose \cite{P1} for detailed presentation of this viewpoint. 

Therefore the following statement of Chaitin, {\it ``the manifest of computability''},\index{manifest of computability}  is not about science done by humans, but science done 
by computers or other artificial intellectual systems, so see \cite{Chaitin1}, p. 64:  

{\small ``I think of a scientific theory as a binary computer program  for calculating the observations, which are also written in binary. And you have a law of nature if there 
is compression, if the experimental data is compressed in a computer program that has a smaller number of bits than are in the data that it explains. ...

But if the experimental data  cannot be compressed, if the smallest program for calculating it is just as large as it is ..., then the data is lawless, unstructured, patternless, not 
amenable to scientific study, incompressible. In a word, random, irreducible!''} 

Now we turn to quantum randomness. As was pointed out, the mathematical theory of randomness cannot distinguish ``classical and quantum randomness'',
random sequences generated by coin tossing and by quantum random generators. They are equally algorithmically complex (Kolmogorov) and typical (Martin-L\"of).\footnote{
We remind that ``a good theory of randomness as unpredictability'', a la von Mises, has not been yet created. Its development culminated in Wald's theorem, section \ref{randomness}.
The next step, to the Church-Wald collectives,  might be a step in wrong direction.} How can one try to formalize the notion of quantum random sequence? Combing the viewpoints of 
von Neumann  and Kolmogorov-Martin-L\"of  (and Church, Solomonoff, Chaitin, Schnorr), we 
can say that this is  a Kolmogorov-Martin-L\"of  random sequence such that it is impossible to present a causal model of its generation.  
(A  larger class of quantum randomness one gets by considering a deterministic dynamical system, instead of a general causal model.) 

However, it seems to  be impossible to prove the impossibility of causal generation for a concrete sequence. In spite of huge activity in generation of various no-go theorems, we  still 
do not have an adequate no-go theorem for one output  experiment with quantum systems. 
The famous Bell theorem is about impossibility of combination of a few causally generated outputs (if we ignore the issue of nonlocality).\footnote{Bell's inequality contains statistical 
data  collected in a few incompatible experimental tests, see \cite{KHR_CONT}.} This theorem
 cannot exclude the possibility that each of them can be causally generated. It even cannot exclude the possibility of
deterministic generation of all these outputs. An adequate no-go theorem might be the original von Neumann theorem \cite{VN},
 see also section \ref{VN_RD}. However, nowadays it is commonly considered
as inadequate  to the real quantum mechanical situation.

Finally, we remark that one has to distinguish   quantum randomness  and quantum probability. It seems that these notions are often identified (may be unconsciously). Then the nonclassical structure 
of quantum probability is treated as the argument in favor of nonclassicality of quantum randomness. For example, we can point to intensive studies justifying peculiarities of quantum 
randomness as compared to classical randomness by using the Bell no-go theorem \cite{77}. Here nonclassical probability structure of the Bell test is treated as leading to generation of  
nonclassically random sequences.  

I would like to thank A. Zeilinger, C. Brukner, K. Svozil for critical discussions on the information interpretation and hospitality during my visits to Vienna.
This study was supported by the grant ``Mathematical Modeling of Complex Hierarchic Systems''.


\begin{thebibliography}{199}

\bibitem{BOOK} Asano, M., Khrennikov, A., Ohya, M., Tanaka, Y. and Yamato, I. (2015). 
\emph{ Quantum Adaptivity in Biology: from Genetics to Cognition} (Springer, Heidelberg-Berlin-New York).

\bibitem{BIO} M. Asano, I. Basieva, A. Khrennikov, M. Ohya, Y. Tanaka, I. Yamato, Quantum information biology: from information interpretation of quantum mechanics
 to applications in molecular  biology and cognitive psychology. {\it Found.  Phys.} {\bf 45},  N 10,  1362-1378 (2015). 

\bibitem{[13]} Ballentine, L.~E. (1970). The statistical interpretation of quantum mechanics, \emph{ Rev. Mod. Phys.} {\bf 42}, pp. 358--381. 

\bibitem{[15]} Ballentine, L.~E. (1989). \emph{ Quantum mechanics} (Englewood Cliffs,  New Jersey). 

\bibitem{B0} Bell, J.~S. (1964). On the Einstein-Podolsky-Rosen Paradox, \emph{Physics 1} {\bf 195}.

\bibitem{B7} Bell, J.~S. (1982). On the impossible pilot wave, \emph{Ref.TH.3315-CERN}, p. 15.

\bibitem{B} Bell, J. (1987). \emph{Speakable and Unspeakable in Quantum Mechanics} (Cambridge Univ. Press, Cambridge). 

\bibitem{[26]}  Bohm, D.  and Hiley, B.  (1993). \emph{ The undivided universe: an ontological interpretation of quantum mechanics} (Routledge and Kegan Paul, London).

 \bibitem{BR1}    Brukner, C.  and Zeilinger, A. (1999). Malus' law and quantum information, \emph{Acta Physica Slovava} {\bf 49} 4, pp. 647--652.

 \bibitem{BR2}    Brukner, C.  and Zeilinger, A. (1999). Operationally invariant information in quantum mechanics, \emph{Phys. Rev. Lett.} {\bf 83} 17, pp. 3354--3357.
 
\bibitem{BR3} Brukner, C.  and Zeilinger, A. (2009). Information Invariance and Quantum Probabilities, \emph{Found. Phys.} {\bf 39}, p. 677.

\bibitem{BR4} Brukner, C.,   \emph{On the quantum measurement problem}, Preprint	arXiv:1507.05255 [quant-ph].

\bibitem{Chaitin} Chaitin, G.~J. (1969). On the Simplicity and Speed of Programs for Computing Infinite Sets 
of Natural Numbers,  \emph{Journal of the ACM} {\bf 16}, pp. 407--422.

\bibitem{Chaitin1} Chaitin, G.~J. (2005). \emph{Meta Math!} (Pantheon Books, New York). 

\bibitem{Caves1}  Caves, C.~M., Fuchs, C.~A. and Schack, R. (2002). Quantum probabilities as Bayesian probabilities, \emph{Phys. Rev. A} {\bf 65}, 022305.

\bibitem{CH} Church, A. (1940). On the concept of a random sequence, \emph{Bull. Amer. Math. Soc.} {\bf 46}, pp. 130--135.

\bibitem{DBB}  De Broglie, L. (1964). \emph{The Current Interpretation of Wave Mechanics: a Critical Study}
(with a chapter by A.~E. Silva), (Elsevier Publishing Co).  

\bibitem{Raedt1} De Raedt, K., Keimpema,  K., De Raedt,   H., Michielsen,   K.  and Miyashita, S. (2006). A local realist model for correlations of the singlet state, \emph{The European Physical Journal B} {\bf 53}, pp. 139--142. 

\bibitem{CT2}  Diaconis, P., Holmes, S. and Montgomery, R. (2007). Dynamical bias in the coin toss, \emph{SIAM Rev.} {\bf 49}, p. 211.

\bibitem{Feller} Feller, W. (1968). \emph{An Introduction to Probability Theory and its Applications}, Vol.~1 (John Wiley \& Sons, New York).

\bibitem{Fuchs1} Fuchs, C.~A. (2002). Quantum mechanics as quantum information (and only a little more), in A. Khrennikov (ed.), \emph{Quantum Theory: Reconsideration
of Foundations, Ser. Math. Modeling} {\bf 2} (V\"axj\"o University Press, V\"axj\"o), pp. 463--543. 

\bibitem{Fuchs2a} Fuchs, C. (2007). Delirium quantum (or, where I will take quantum mechanics if it will let me), in G. Adenier, C. Fuchs and A.~Yu. Khrennikov (eds.), 
Foundations of Probability and Physics-3, \emph{Ser. Conference Proceedings} {\bf 889} (American Institute of Physics,  Melville, NY), pp. 438--462.  

\bibitem{Fuchs3} Fuchs, C.~A. and Schack, R. (2011). A Quantum-Bayesian Route to Quantum-State Space, \emph{Found. Phys.} {\bf 41}, p. 345.

\bibitem{Fuchs6} Fuchs, C.~A.,  Mermin, N.~D. and Schack, R. (2014). An Introduction to QBism with an Application to the Locality of Quantum Mechanics, \emph{Am. J. Phys.} {\bf 82}, p. 749.

\bibitem{Giustina} M. Giustina et al.,  (2015). Significant-loophole-free test of Bell's theorem with entangled photons. \emph{ Phys. Rev. Lett.} 115, 250401.

\bibitem{Hensen}  B. Hensen et al. (2015). Loophole-free Bell inequality violation using electron spins separated by 1.3 Kilometres. \emph{Nature} {\bf 526}, 682.

\bibitem{Hess1} Hess, K. and Philipp, W. (2005). Bell's theorem: critique of proofs with and without inequalities, in G. Adenier, A.~Yu.  Khrennikov (eds.), 
Foundations of Probability and Physics-3, \emph{Ser. Conference Proceedings} {\bf 750} (American Institute of Physics,  Melville, NY), 
pp. 150--155. 

\bibitem{Hess2} Hess, K. (2007). In memoriam Walter Philipp, in G. Adenier, C. Fuchs and A.~Yu. Khrennikov (eds.),
Foundations of Probability and Physics-3, \emph{Ser. Conference Proceedings} {\bf 889} (American Institute of Physics, Melville, NY), pp. 3--6. 

\bibitem{INT_KHR} (1999). \emph{ Interpretations of Probability} (VSP Int. Sc. Publishers, Utrecht/Tokyo). Secodn edition:   
Khrennikov, A. (2009). \emph{Interpretations of Probability}, 2nd edn. (De Gruyter, Berlin).

\bibitem{KHR_CONT} Khrennikov,  A. ( 2009). {\it Contextual Approach to Quantum Formalism,} (Springer, Berlin-Heidelberg-New York)

\bibitem{QBism_KHR} Khrennikov, A. (2015) External observer reflections on QBism. arXiv:1512.07195 [quant-ph] 

\bibitem{UB_KHR} Khrennikov, A. (2010). \emph{Ubiquitous  quantum structure: from psychology to finances} (Springer, Berlin-Heidelberg-New York).

\bibitem{ST2}  Khrennikov, A., Ramelow, S., Ursin, R., Wittmann, B., Kofler, J., Basieva, I. (2014). On the equivalence of the Clauser-Horne and
Eberhard inequality based tests, \emph{Phys. Scr.} {\bf T163}, 014019. 

\bibitem{Kofler}  Kofler, J.and Zeilinger, A. (2010).  Quantum information and randomness, \emph{European Rev.} {\bf 18},  N 4, pp. 469--480.

\bibitem{[134]a}  Kolmolgorov, A.~N. (1963). On Tables of Random Numbers, \emph{Sankhya Ser. A} {\bf 25}, pp.  369--375.

\bibitem{[134]b }  Kolmolgorov, A.~N.  (1998). On Tables of Random Numbers, \emph{Theor. Comp. Sc.} {\bf 207 (2)}, pp. 387--395. 

\bibitem{[134]} Kolmolgorov, A.~N. (1965). Three approaches to the quantitative definition of information, \emph{Problems Inform. Transmition} {\bf 1}, pp. 1--7.

\bibitem{[135]} Kolmolgorov, A.~N. (1968). Logical basis for information theory and probability theory, \emph{IEEE Trans. IT} {\bf 14}, pp. 662--664.


\bibitem{KP1} Kupczynski, M. (2007). EPR paradox, locality and completeness of quantum theory, in G.
Adenier, A.   Khrennikov, P. Lahti, V. Man'ko and T. Nieuwenhuizen (eds.), \emph{Quantum Theory Reconsideration of Foundations-4, AIP Conference Proceedings} {\bf 962} (American Institute of Physics, Melville, NY), pp. 274--285.

\bibitem{KP2} Kupczynski, M. (2011). Time Series, Stochastic Processes and Completeness of Quantum Theory Advances Quantum Theory, in 
G. Jaeger, A. Khrennikov, M. Schlosshauer, G. Weihs (eds.),  \emph{AIP Conference Proceedings} {\bf 1327} (American Institute of Physics, Melville, NY), pp. 394--400.

\bibitem{Laplace}  de Laplace, P.-S. (1819, 1952). \emph{A Philosphical Essay on Probabilities} (Dover), translated from 6th french edn.




\bibitem{[142]} Martin-L\"of, P. (1966). On the concept of random sequence, \emph{Theory of Probability Appl.} {\bf 11}, pp. 177--179.

\bibitem{[143]} Martin-L\"of, P. (1966). The definition of random sequence, \emph{Inform. Contr.} {\bf 9}, pp. 602--619.

\bibitem{[144]} Martin-L\"of, P. (1971). Complexity oscillations in infinite binary sequences, \emph{Z. Wahrscheinlichkeitstheorie verw.} {\bf 19}, pp. 225--230.

\bibitem{MIH}  Michielsen, K.,  De Raedt, H. and  Hess, K. (2011). Boole  and Bell inequality, in G. Jaeger, A. Khrennikov, M. Schlosshauer, G. Weihs (eds.), \emph{Advances in Quantum Theory, AIP Conference Proceedings}, Vol.~ 1327 (Melville--New York), pp. 429--433. 

\bibitem{CT3} Mizuguchi, T. and Suwashita, M. (2006). Dynamics of coin tossing, \emph{Prog. Theor. Phys. Suppl} {\bf 161}, p. 274.

\bibitem{Theo}  Nieuwenhuizen, Th.~M. (2009). Where Bell went wrong? in L. Accardi, G. Adenier, C.~A. Fuchs, 
G. Jaeger, A.~Yu. Khrennikov, J.-A. Larsson and S. Stenholm (eds.), \emph{AIP Conf. Proc., Foundations of Probability and Physics - 5} {\bf 1101} (Am. Inst. Phys., Melville, NY), pp. 127--133. 

\bibitem{P1} Penrose, R. (1989). \emph{The Emperor's new mind} (Oxford Univ. Press, New-York). 

\bibitem{77}  Pironio, S.,  Acin, A.,  Massar, S.,  Boyer de la Giroday, A.,  Matsukevich, D.~N.,  Maunz, P.,  Olmschenk, S.,  Hayes, D.,  Luo, L.,  Manning, T.~A. and  Monroe, C. (2010). 
Random Numbers Certified by Bell's Theorem, 	\emph{Nature} {\bf 464}, p. 1021.

\bibitem{[158]} Schnorr, C.~P.  (1971). Zufalligkeit und Wahrscheinlichkeit, \emph{Lect. Notes in Math.} {\bf 218} (Springer-Verlag, Berlin).

\bibitem{Shalm} L. K. Shalm et al. (2015). Strong loophole-free test of local realism.  \emph{Phys. Rev. Lett.} {\bf 115}, 250402.


\bibitem{Sol0}  Solomonoff, R.~J. (1960). A preliminary report on a general theory of inductive inference, \emph{Report V-131} (Cambridge, Ma.: Zator Co.).

\bibitem{Sol}   Solomonoff, R.~J. (1964). A formal theory of inductive inference, \emph{Information and Control} {\bf 7}, pp. 1--22.


\bibitem{COIN}  Strzalko, J.,  Grabski, J., Stefanski, A., Perlikowski, P. and Kapitaniak, T. (2010).  Understanding coin-tossing, \emph{Math. Intelligence} {\bf  32}, N 4, pp.  54--58.

\bibitem{Ville} Ville, J. (1939). \emph{Etude critique de la notion de collective} (Gauthier-Villars, Paris).

\bibitem{VN} von Neuman, J. (1955). \emph{Mathematical foundations of quantum mechanics} (Princeton Univ. Press, Princenton).

\bibitem{[169]} von Mises, R. (1919). Grundlagen der Wahrscheinlichkeitsrechnung, \emph{Math. Z.} {\bf 5}, pp. 52--99.

\bibitem{[170]} von Mises, R. (1957). \emph{Probability, Statistics and Truth} (Macmillan, London).

\bibitem{[171]} von Mises, R. (1964). \emph{The mathematical theory of probability and statistics} (Academic, London).

\bibitem{Ville} Ville, J. (1939). \emph{Etude critique de la notion de collective} (Gauthier-Villars, Paris).

\bibitem{CT1}   Vulovic, V.~Z. and  Prange, R.~E. (1986). Randomness of true coin toss, \emph{Phys. Rev. A} {\bf 33}, p. 576.


\bibitem{[173]} Wald, A. (1938). Die Widerspruchsfreiheit des Kollektivbegriffs in der Wahrscheinlichkeitsrechnung, \emph{Ergebnisse eines Math. Kolloquiums} {\bf 8}, pp. 38--72.


\bibitem{Wheeler} Wheeler, John A. (1990). ``Information, physics, quantum: The search for links'', in Zurek, Wojciech Hubert \emph{Complexity, Entropy, and the Physics of Information} (Redwood City, California: Addison-Wesley).

\bibitem{ST1}  Zhang, Ya.,  Glancy, S. and  Knill, E. (2011). Asymptotically optimal data analysis for rejecting local realism, \emph{Phys. Rev. A} {\bf 84}, 062118.

\bibitem{Z0} Zeilinger, A.  (1999). A foundational principle for quantum mechanics, \emph{Foundations of Physics} {\bf 29}, 6pp. 31--641.

\bibitem{Z1} Zeilinger, A. (2010).  \emph{Dance of the Photons: From Einstein to Quantum Teleportation} (Farrar, Straus and Giroux, New-York). 
  
\end{thebibliography}
\end{document}